\documentclass[twocolumn,aps,prd,epsf,showpacs]{revtex4-1}
\pdfoutput=1
\usepackage[pdftex]{graphicx}
\usepackage{color}
\usepackage{amsmath}
\usepackage{amsfonts}
\usepackage{amssymb}
\usepackage{dsfont}
\usepackage[]{units}
\usepackage{braket}
\usepackage[utf8]{inputenc}
\usepackage[T1]{fontenc}
\usepackage{url}
\usepackage[english]{babel}
\usepackage{bm}
\usepackage{verbatim}
\usepackage{textcomp}

\newcommand{\be}{\begin{equation}}
\newcommand{\ee}{\end{equation}}
\newcommand{\bea}{\begin{eqnarray}}
\newcommand{\eea}{\end{eqnarray}}

\newcommand{\bei}{\begin{itemize}}

\newcommand{\eei}{\end{itemize}}
\newcommand{\ben}{\begin{enumerate}}
\newcommand{\een}{\end{enumerate}}

\newcommand{\gtapprox}{\raisebox{-0.5ex}{$\,\stackrel{>}{\scriptstyle\sim}\,$}}
\newcommand{\ltapprox}{\raisebox{-0.5ex}{$\,\stackrel{<}{\scriptstyle\sim}\,$}}

\begin{document}

\title{ 
$u d \bar{b} \bar{b}$ tetraquark resonances with lattice QCD potentials \\ and the Born-Oppenheimer approximation
}

\author{$^{(1)}$Pedro Bicudo}
\email{bicudo@tecnico.ulisboa.pt}

\author{$^{(1)}$Marco Cardoso}
\email{mjdcc@cftp.ist.utl.pt}

\author{$^{(2)}$Antje Peters}
\email{peters@th.physik.uni-frankfurt.de}

\author{$^{(2)}$Martin Pflaumer}
\email{pflaumer@th.physik.uni-frankfurt.de}

\author{$^{(2)}$Marc Wagner}
\email{mwagner@th.physik.uni-frankfurt.de}

\affiliation{\vspace{0.1cm}$^{(1)}$CFTP, Dep.\ F\'{\i}sica, Instituto Superior T\'ecnico, Universidade de Lisboa, Av.\ Rovisco Pais, 1049-001 Lisboa, Portugal}

\affiliation{\vspace{0.1cm}$^{(2)}$Johann Wolfgang Goethe-Universit\"at Frankfurt am Main, Institut f\"ur Theoretische Physik, Max-von-Laue-Stra{\ss}e 1, D-60438 Frankfurt am Main, Germany}

\begin{abstract}
We study tetraquark resonances with lattice QCD potentials computed for a static $\bar{b} \bar{b}$ pair in the presence of two lighter quarks $u d$, the Born-Oppenheimer approximation and the emergent wave method. As a proof of concept we focus on the system with isospin $I = 0$, but consider different relative angular momenta $l$ of the heavy quarks $\bar{b} \bar{b}$. For $l=0$ a bound state has already been predicted with quantum numbers $I(J^P) = 0(1^+)$. Exploring various angular momenta we now compute the phase shifts and search for $\mbox{S}$ and $\mbox{T}$ matrix poles in the second Riemann sheet. We predict a tetraquark resonance for $l =1$, decaying into two $B$ mesons, with quantum numbers $I(J^P) = 0(1^-)$, mass $m = 10 \, 576^{+4}_{-4} \, \textrm{MeV}$ and decay width $\Gamma = 112^{+90}_{-103} \, \textrm{MeV}$.
\end{abstract}


\maketitle


\section{Introduction \label{sec:intro}}

A long standing problem in particle physics is to understand exotic hadrons, i.e.\ hadrons which have a structure more complicated than a quark-antiquark pair or a triplet of quarks \cite{Jaffe:1976ig}. The problem of identifying exotic hadrons, say tetraquarks, pentaquarks, hexaquarks, hybrids or glueballs -- expected since the onset of QCD -- turned out to be much harder than initially expected \cite{Bicudo:2015bra}. The observed candidates are resonances high in the spectrum, not only difficult to observe, but also technical to address in quark or hadron models. They possibly require the development of new techniques, potentially relevant to other areas of physics, to be studied theoretically from first principles, e.g.\ with lattice QCD \cite{Wu:2014vma,Morningstar:2016arm}.

Our main motivation is to investigate tetraquarks by combining lattice QCD and quantum mechanics techniques. We specialize in systems with two heavy antiquarks, which are expected to form bound states, when sufficiently heavy \cite{Ader:1981db,Ballot:1983iv,Heller:1986bt,Carlson:1987hh,Lipkin:1986dw,Brink:1998as,Gelman:2002wf,Vijande:2003ki,Janc:2004qn,Cohen:2006jg,Vijande:2007ix}. The starting point are potentials of two static antiquarks in the presence of two light quarks, which can be computed with state of the art lattice QCD techniques (cf.\ e.g.\ \cite{Detmold:2007wk,Wagner:2010ad,Bali:2010xa,Wagner:2011ev,Brown:2012tm,Bicudo:2015kna}). If the masses of the two heavy quarks are much larger than the scale of QCD, which is the case for two $\bar{b}$ quarks, their dynamics can then be described by a quantum mechanical Hamiltonian with the aforementioned lattice QCD potentials. This two-step approach is the Born-Oppenheimer approximation \cite{Born:1927}. Using this approach, a $u d \bar{b} \bar{b}$ tetraquark bound state with quantum numbers $I(J^P) = 0(1^+)$ has recently been predicted \cite{Bicudo:2012qt,Brown:2012tm,Bicudo:2015vta,Bicudo:2015kna,Bicudo:2016ooe} and confirmed by a lattice QCD computation with four quarks of finite mass \cite{Francis:2016hui}. So far, however, resonances have not been studied in this framework.

Notice there are two classes of double-heavy tetraquarks. The tetraquarks with one heavy quark and one heavy antiquark including the $Z_c$ and $Z_b$ are easier to detect experimentally. Their observation at Belle \cite{Belle:2011aa,Liu:2013dau,Chilikin:2014bkk}, Cleo-C \cite{Xiao:2013iha},
BESIII \cite{Ablikim:2013mio,Ablikim:2013emm,Ablikim:2013wzq,Ablikim:2013xfr,Ablikim:2014dxl} and LHCb
\cite{Aaij:2014jqa} collaborations turned tetraquarks into a main highlight of particle physics in recent years. But since they have more coupled channels we opt here to study tetraquarks with two heavy antiquarks (or quarks), which are theoretically simpler. This ``theoretical simplicity'' is convenient for a first study of resonances with lattice QCD potentials. Moreover, with the recent observation at LHCb of hadronic systems with two heavy quarks \cite{Maciula:2016wci,Aaij:2017ueg} we expect this second class of tetraquarks to be observed in the near future.

In this work we extent the previous Born-Oppenheimer studies with lattice QCD potentials, reviewed in Section~\ref{sec:lattice}. We utilize the emergent wave method, a technique from scattering theory detailed in Section~\ref{sec:emergent}, to compute phase shifts, $\mbox{S}$ and $\mbox{T}$ matrix poles in the second Riemann sheet and the corresponding resonance masses and decay widths. For the first time, we apply this technique with lattice QCD potentials, and our results are presented in Section~\ref{sec:results}. We conclude in Section~\ref{sec:conclusion}.


\section{Lattice QCD potentials of two static antiquarks in the presence of two light quarks and prediction of a stable $u d \bar{b} \bar{b}$ tetraquark\label{sec:lattice}}

In preceding papers we have computed potentials $V(r)$ of two static antiquarks $\bar{Q} \bar{Q}$ in the presence of two light quarks $q q$ using lattice QCD. The computations have been carried out for many different quantum numbers including light flavor combinations $q q$ with $q \in \{ u , d, s, c \}$, parity $P$ and total angular momentum of the light quarks and gluons $j$ (cf.\ e.g.\ \cite{Bicudo:2015vta,Bicudo:2015kna}). There are both attractive and repulsive channels. Most promising with respect to the existence of tetraquark bound states or resonances are attractive potentials with light quarks $q \in \{ u , d\}$, since they are rather wide and deep. There are two such potentials, with quantum numbers $(I = 0,j = 0)$ and $(I = 1,j = 1)$, where $I$ denotes isospin.

\begin{figure}[htb]
\centerline{\includegraphics[width=0.95\columnwidth]{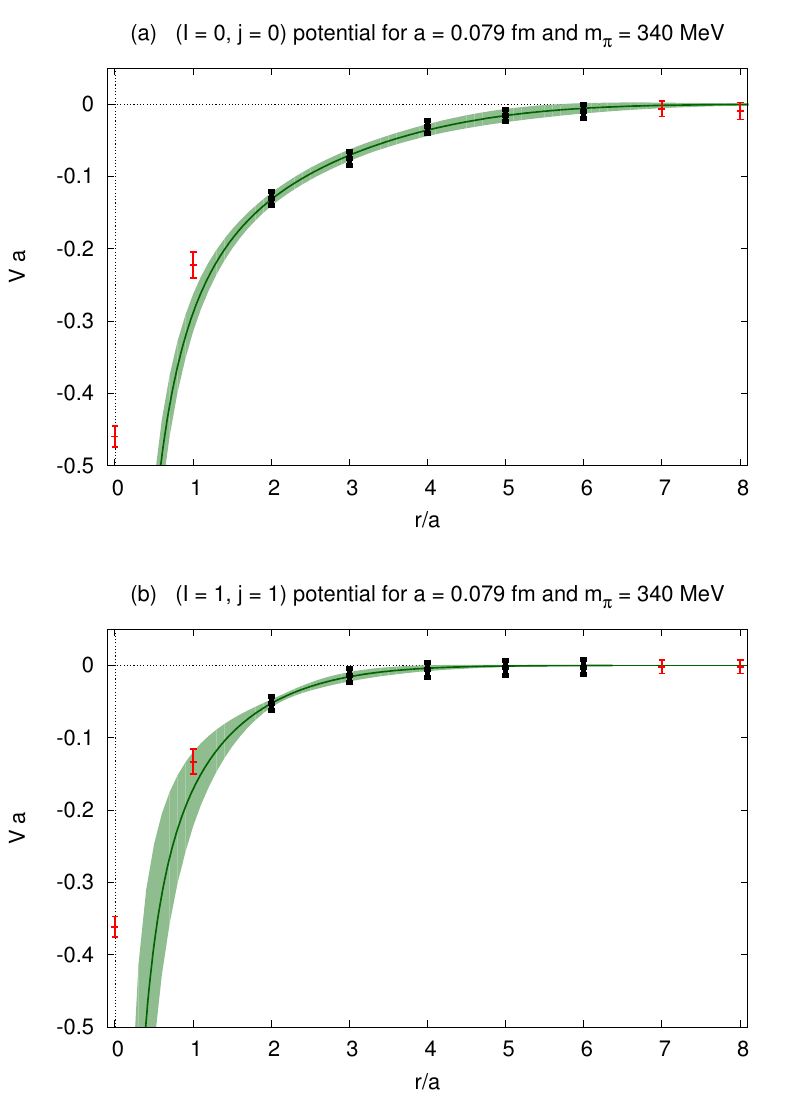}}
\caption{\label{FIG_potentials}(Colour online.) (a)~$(I = 0,j = 0)$ potential. (b)~$(I = 1,j = 1)$ potential.}
\end{figure}

We have used creation operators
\begin{eqnarray}
\nonumber & & \hspace{-0.7cm} \mathcal{O}[f,f',\Gamma,\tilde{\Gamma}] = (\mathcal{C} \Gamma)_{A B} (\mathcal{C} \tilde{\Gamma})_{C D} \\
 & & = \Big(\bar{Q}_C^a(\mathbf{r}_1) \psi_A^{(f) a}(\mathbf{r}_1)\Big) \Big(\bar{Q}_D^b(\mathbf{r}_2) \psi_B^{(f') b}(\mathbf{r}_2)\Big) \ ,
\end{eqnarray}
where $r = |\mathbf{r}_2 - \mathbf{r}_1|$, $a,b$ denote color and $A,B,C,D$ spin indices and $\psi^{(f)} \psi^{(f')} = u d - d u$ for $I = 0$ and $\psi^{(f)} \psi^{(f')} \in \{ uu ,  u d + d u , dd \}$ for $I = 1$. For the $(I = 0,j = 0)$ potential $\Gamma = (1 + \gamma_0) \gamma_5$, while for the $(I = 1,j = 1)$ potential $\Gamma = (1 + \gamma_0) \gamma_j$ ($j = 1,2,3$). Since the potentials are independent of the static quark spins, one can choose arbitrarily $\tilde{\Gamma} \in \{ (1 - \gamma_0) \gamma_5 , (1 - \gamma_0) \gamma_j \}$. As usual in lattice QCD hadron spectroscopy we have extracted the potentials from the asymptotic exponential decay in the temporal separation $t$ of correlation functions
\begin{equation}
\langle \Omega | \mathcal{O}[f,f',\Gamma,\tilde{\Gamma}]^\dagger(t) \mathcal{O}[f,f',\Gamma,\tilde{\Gamma}](0) | \Omega \rangle \ .
\label{eq:corr_fnc} 
\end{equation}
Example plots for lattice spacing $a \approx 0.079 \, \textrm{fm}$ and $u/d$ quark masses corresponding to a pion mass $m_\pi \approx 340 \, \textrm{MeV}$ are shown in Fig.\ \ref{FIG_potentials}.

Since it is known that the existence of a stable tetraquark as well as its binding energy exhibits a sizable dependence on the light quark mass \cite{Bicudo:2015vta}, we have performed computations of the potentials for three different $u/d$ quark masses corresponding to $m_\pi \in \{ 340 \, \textrm{MeV} , 480 \, \textrm{MeV} , 650 \, \textrm{MeV} \}$. Then we have used these results to extrapolate to the physical $u/d$ quark mass corresponding to $m_\pi = 140 \, \textrm{MeV}$. Moreover, we have crudely estimated systematic errors due to the finite lattice spacing $a \approx 0.079 \, \textrm{fm}$ by performing the computations with two different Wilson twisted mass lattice QCD discretizations. We have found that discretization errors are negligible compared to the current statistical uncertainties (for more details cf.\ \cite{Bicudo:2015kna}). Similarly, effects due to the finite spatial volume of the lattice are expected to be negligible as well.

\begin{figure}[htb]
\centerline{%
\includegraphics[width=0.95\columnwidth]{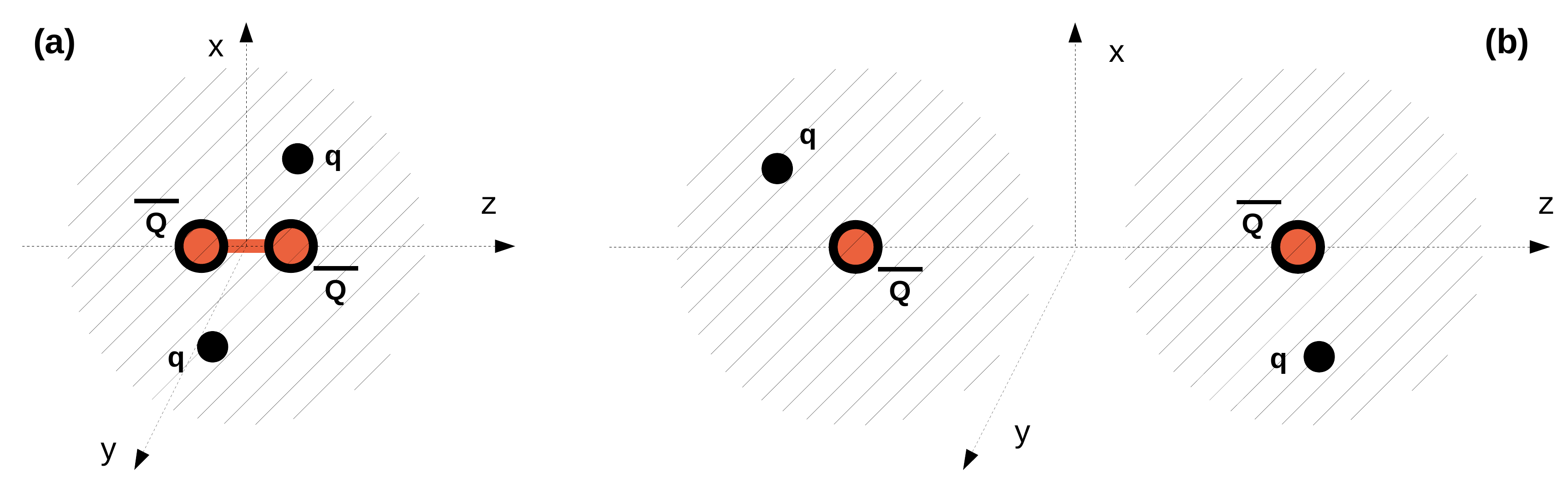}}
\caption{(Colour online.) (a)~At small separations the static antiquarks $\bar{Q} \bar{Q}$ interact by perturbative one-gluon exchange. (b)~At large separations the light quarks $q q$ screen the interaction and the four quarks form two rather weakly interacting $B$ mesons.
\label{fig:screening}}
\end{figure}

To search for bound states and resonances we parameterize the potentials by a screened Coulomb potential,
\begin{equation}
V (r) = -\frac{\alpha}{r} e^{-r^2 / d^2} +V_0 \ .
\label{eq:potential}
\end{equation}
This ansatz is inspired by one-gluon exchange at small $\bar{Q} \bar{Q}$ separations $r$ and a screening of the Coulomb potential due to the formation of two $B$ mesons at large $r$, as illustrated in Fig.\ \ref{fig:screening}. The ansatz, even though phenomenologically motivated, is consistent with our lattice QCD results, which are based on first principles, i.e.\ a fit of (\ref{eq:potential}) to the lattice QCD data yields an acceptable $\chi^2/\textrm{dof} \ltapprox 1$. Vice versa, parameterizing the lattice QCD data by using ans\"atze different from (\ref{eq:potential}) leads e.g.\ to similar results for masses of tetraquark bound states. The values of the two parameters $\alpha$ and $d$ as determined in \cite{Bicudo:2015kna} are listed in Table \ref{tab:parameters}. Clearly, the $(I = 0,j = 0)$ potential is more attractive than the $(I = 1,j = 1)$ potential. Note that there is also an uncertainty associated with the lattice spacing, $a = 0.079(3) \, \textrm{fm}$ (cf.\ \cite{Baron:2009wt}), which is not included in the parameter $d$ in Table~\ref{tab:parameters}. We investigate the effect of this uncertainty at the end of our analysis in section~\ref{sec:results}.

\begin{table}[htb]
\begin{tabular}{cc|cc}
\hline
& & & \vspace{-0.4cm} \\
$I$ & $j$ & $\alpha$ & $d$ in $\textrm{fm}$ \\
& & & \vspace{-0.4cm} \\
\hline
& & & \vspace{-0.3cm} \\
$\ 0 \ $ & $\ 0 \ $ & $\ 0.34^{+0.03}_{-0.03} \ $ & $\ 0.45^{+0.12}_{-0.10} \ $ \\
& & & \vspace{-0.25cm} \\
$1$ & $1$ & $0.29^{+0.05}_{-0.06}$ & $0.16^{+0.05}_{-0.02}$\vspace{-0.3cm} \\
& & & \\
\hline
\end{tabular}
\caption{Parameters $\alpha$ and $d$ of the potential of Eq.\ (\ref{eq:potential}) for two static antiquarks $\bar{Q} \bar{Q}$, in the presence of two light quarks $q q$ with quantum numbers $I$ and $j$, as determined in \cite{Bicudo:2015kna}.
\label{tab:parameters}}
\end{table}

Finally we have applied the Born-Oppenheimer approximation, where Eq.\ (\ref{eq:potential}) is used as a potential for two heavy antiquarks, i.e.\ $\bar{b} \bar{b}$, in the presence of two light quarks $u d$ or for two heavy-light mesons, i.e.\ $B^{(\ast)} B^{(\ast)}$. Solving the Schr\"odinger equation for the $(I = 0,j = 0)$ potential and angular momentum $l = 0$ of the two $\bar{b}$ quarks a bound state has been predicted with binding energy $90^{+43}_{-36} \textrm{MeV}$ and quantum numbers $I(J^P) = 0(1^+)$ \cite{Bicudo:2015kna}.

The use of the Born-Oppenheimer approximation entails a systematic error from quantizing the $\bar{b} \bar{b}$ system with the kinetic energy only. The kinetic energy naturally emerges in the next to leading term in a non-relativistic series expansion. However, the spin dependent terms of the potential are of the same non-relativistic expansion order of the kinetic energy and so far we have not taken them into account directly. Nevertheless, in Ref.\ \cite{Bicudo:2016ooe} the spin effects have been estimated and they have little effect on the binding energy of the tetraquark. Finally, very recent computations in lattice QCD with non-relativistic bottom quarks, which account for both the kinetic and spin effects, confirm our previous results for the binding energy, obtained with the Born-Oppenheimer approximation \cite{Junnarkar:2017lat}. Thus we expect that the use of the Born-Oppenheimer approximation is  adequate for our study.


\section{The emergent wave method \label{sec:emergent}}

We now summarize the emergent wave method, explained in detail for instance in Ref.\ \cite{Bicudo:2015bra}, which is suited to study phase shifts and resonances. Let us consider the same Schr\"odinger equation utilized in the bound state study,
\begin{equation}
\Big(H_{0} + V(r)\Big) \Psi = E \Psi \ .
\label{eq:schro}
\end{equation}
The first step is to split the wave function into two parts,
\begin{equation}
\Psi = \Psi_{0} + X \ ,
\label{eq:sep_psi}
\end{equation}
where $\Psi_{0}$ is the incident wave, a solution of the free Schr\"odinger equation,
\begin{equation}
H_{0} \Psi_{0} = E \Psi_{0} ,
\label{eq:schro_free}
\end{equation}
and $X$ is the emergent wave. Substituting Eq. (\ref{eq:sep_psi}) into Eq. (\ref{eq:schro}) and using Eq. (\ref{eq:schro_free}) we obtain
\begin{equation}
\Big(H_{0} + V(r) - E\Big) X = -V(r) \Psi_{0} \ .
\label{eq:schro_scatter}
\end{equation}
For any energy $E$ we can use this equation to calculate the emergent wave $X$ by providing the corresponding
$\Psi_{0}$ and fixing the appropriate boundary conditions. From the asymptotic behaviour of $X$ we then determine the phase shifts, the $\mbox{S}$ matrix and the $\mbox{T}$ matrix. 

The problem can be continued to complex energies in a straightforward way and we can, therefore, find the poles of the $\mbox{S}$ matrix and the $\mbox{T}$ matrix in the complex plane. We identify a resonance with a pole, when located in the second Riemann sheet at $m - i \Gamma/2$, where $m$ is the mass and $\Gamma$ is the decay width of the resonance.


\subsection{Partial wave decomposition}

The Hamiltonian describing the two heavy antiquarks $\bar{b} \bar{b}$ at vanishing total momentum, i.e.\ in the rest frame of the system, is
\begin{equation}
H = H_0 + V(r) = -\frac{\hbar^{2}}{2 \mu} \triangle + V(r)
\label{EQN005}
\end{equation}
with reduced mass $\mu = M/2$, where $M = 5 \, 280 \, \textrm{MeV}$ is the mass of the $B$ meson from the PDG \cite{Agashe:2014kda}. For simplicity we omit the additive constant $2 M$ in Eq. (\ref{EQN005}), i.e.\ all resulting energy eigenvalues are energy differences with respect to $2 M$. We consider an incident plane wave $\Psi_{0} = e^{i \mathbf{k} \cdot \mathbf{r}}$, which can be expressed as a sum of spherical waves,
\begin{equation}
\Psi_{0} = e^{i \mathbf{k} \cdot \mathbf{r}} = \sum_{l} (2l+1) i^{l} j_{l}(k r) P_{l}(\hat{\mathbf{k}} \cdot \hat{\mathbf{r}}) \ ,
\label{eq:expansionsphericalbessel}
\end{equation}
where $j_{l}$ are spherical Bessel functions, $P_{l}$ are Legendre polynomials and the relation between energy and momentum is $\hbar k = \sqrt{2 \mu E}$. For a spherically symmetric potential $V(r)$ as in Eq. (\ref{eq:potential}) and an incident wave $\Psi_{0} = e^{i \mathbf{k} \cdot \mathbf{r}}$ the emergent wave $X$ can also be expanded in terms of Legendre polynomials $P_{l}$,
\begin{equation}
X = \sum_{l} (2l+1) i^{l} \frac{\chi_l(r)}{k r} P_{l}(\hat{\mathbf{k}} \cdot \hat{\mathbf{r}}) \ .
\label{eq:001}
\end{equation}
Inserting Eq. (\ref{eq:expansionsphericalbessel}) and Eq. (\ref{eq:001}) into Eq. (\ref{eq:schro_scatter}) leads to a set of ordinary differential equations for $\chi_l$,
\begin{eqnarray}
\nonumber & & \hspace{-0.7cm} \bigg(-\frac{\hbar^2}{2 \mu} \frac{d^{2}}{dr^{2}} + \frac{l (l+1)}{2 \mu r^{2}} + V(r) - E\bigg) \chi_l(r) = \\
 & & = -V(r) k r j_l(k r) \ .
\label{eq:1cl0:radial}
\end{eqnarray}


\subsection{Solving the differential equations for the emergent wave}

The potentials $V(r)$, Eq. (\ref{eq:potential}), are exponentially screened, i.e.\ $V(r) \approx 0$ for $r \geq R$, where $R \gg d$. For large separations $r \geq R$ the emergent wave is, hence, a superposition of outgoing spherical waves, i.e.\
\begin{equation}
\frac{\chi_l(r)}{k r} = i \, t_l h_l^{(1)}(k r) ,
\label{eq:002}
\end{equation}
where $h_l^{(1)}$ are the spherical Hankel functions of first kind.

Our aim is now to compute the complex prefactors $t_l$, which will eventually lead to the phase shifts. To this end we solve the ordinary differential equation (\ref{eq:1cl0:radial}). The corresponding boundary conditions are the following:
\begin{itemize}
\item At $r = 0$: $\chi_l(r) \propto r^{l+1}$.

\item For $r \geq R$: Eq. (\ref{eq:002}).
\end{itemize}
Note that the boundary condition for $r \geq R$ depends on $t_l$. For a given value of the energy $E$ this boundary condition is only fulfilled for a specific corresponding value of $t_l$. In other words the boundary condition for $r \geq R$ fixes $t_l$ as a function of $E$.

The numerical solution of the differential Eq.\ (\ref{eq:1cl0:radial}) is rather straightforward. To check our results and to exclude any numerical artefacts we implemented two different approaches: (1) a fine uniform discretization of the interval $[0,R]$, which reduces the differential equation to a large set of linear equations, which can be solved rather efficiently, since the corresponding matrix is tridiagonal; (2) a standard 4-th order Runge-Kutta shooting method.


\subsection{Phase shifts and $\mathbf{S}$ and $\mathbf{T}$ matrix poles}

The quantity $t_l$ is a $\mbox{T}$ matrix eigenvalue (cf.\ standard textbooks on quantum mechanics and scattering, e.g.\ \cite{Merzbacher}). From $t_l$ we can calculate the phase shift $\delta_l$ and also read off the corresponding $\mbox{S}$ matrix eigenvalue $s_l$
\footnote{ At large distances $r \geq R$, the radial wavefunction is $k r [ j_l (kr) + i \, t_l h_l^{(1)}(k r)] =
(k r /2) [ h_l ^{(2)}(kr) +e^{2 i \delta_l} h_l^{(1)}(k r)] $.}
,
\begin{equation}
s_l \equiv 1 + 2 i t_l = e^{2 i \delta_l} \ .
\label{eq:003}
\end{equation}

\begin{figure}[t!]
\centerline{%
\includegraphics[width=0.95\columnwidth]{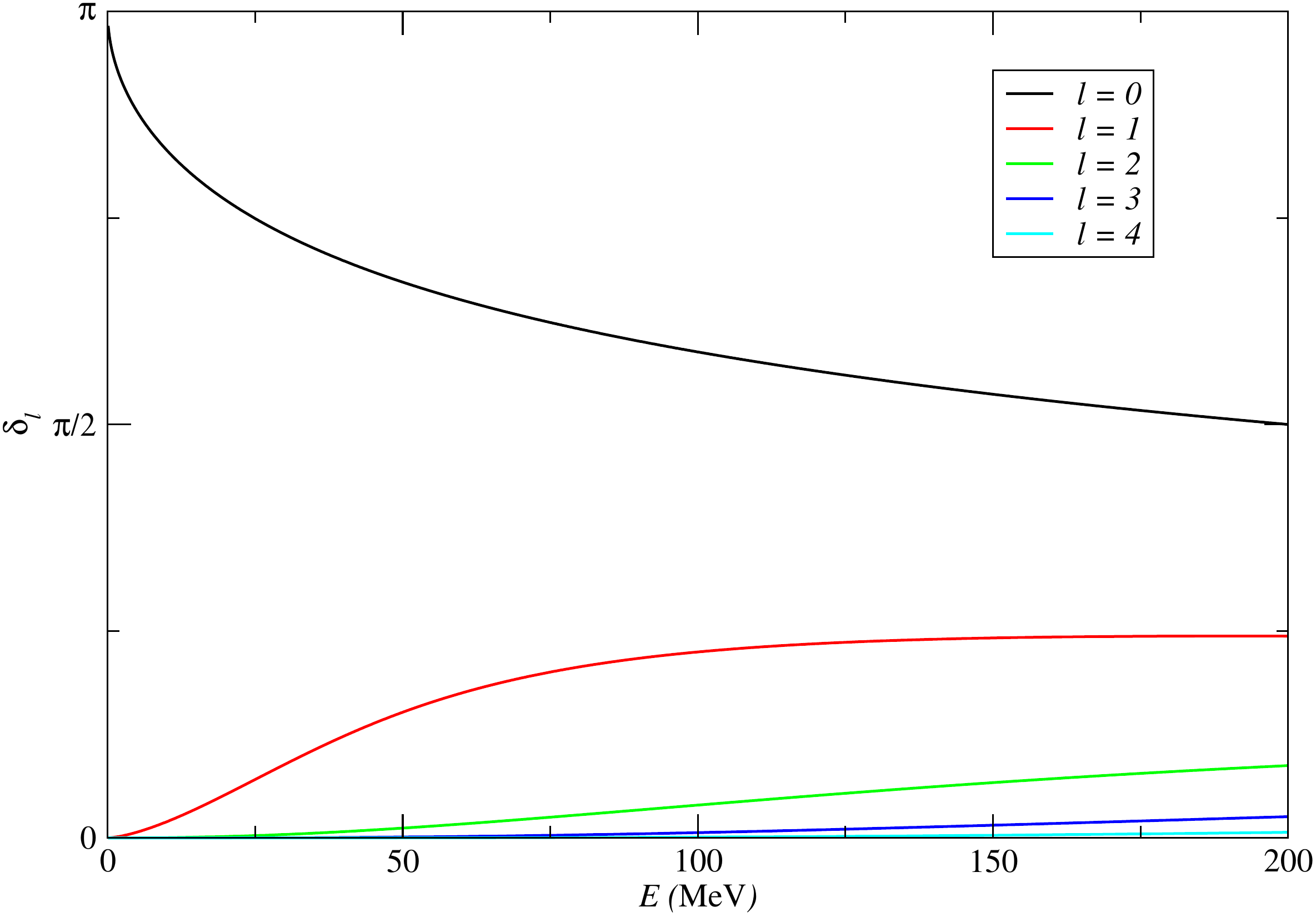}}
\caption{(Colour online.) 
Phase shift $\delta_l$ as a function of the energy $E$ for different angular momenta $l = 0, 1, 2, 3, 4$ for the $(I = 0, j = 0)$ potential ($\alpha = 0.34$, $d = 0.45 \, \textrm{fm}$).
\label{fig:phaseShift_ud_singulett}}
\end{figure}

Moreover, note that both the $\mbox{S}$ matrix and the $\mbox{T}$ matrix are analytical in the complex plane. They are well-defined for complex energies $E$. Thus, our numerical method can as well be applied to solve the differential Eq.\ (\ref{eq:1cl0:radial}) for complex $E$. We find the $\mbox{S}$ and $\mbox{T}$ matrix poles by scanning the complex plane $(\textrm{Re}(E) , \textrm{Im}(E))$ and applying Newton's method to find the roots of $1 / t_l(E)$. The poles of the $\mbox{S}$ and the $\mbox{T}$ matrix correspond to complex energies of resonances. Note the resonance poles must be in the second Riemann sheet with a negative imaginary part both for the energy $E$ and the momentum $k$.


\section{Results for phase shifts, $\mathbf{S}$ matrix and $\mathbf{T}$ matrix poles and resonances \label{sec:results}}

We first consider the more attractive $u d \bar{b} \bar{b}$ potential corresponding to isospin $I = 0$ and light spin $j = 0$ (cf.\ Sec.\ \ref{sec:lattice}). We compute $t_l$ and via Eq. (\ref{eq:003}) the phase shift $\delta_l$ for real energy $E$ and angular momenta $l = 0, 1, 2, \ldots$ A very clear signal for a resonance would be a fast increase of the phase shift $\delta_l$ as a function of $E$ from $0$ to $\approx \pi$, almost like a step function. However, we do not find such a pronounced increase (cf.\ Fig.\ \ref{fig:phaseShift_ud_singulett}). Thus, we must search more thoroughly for possibly existing resonances.

\begin{figure}[t!]
\centerline{%
\includegraphics[width=0.95\columnwidth]{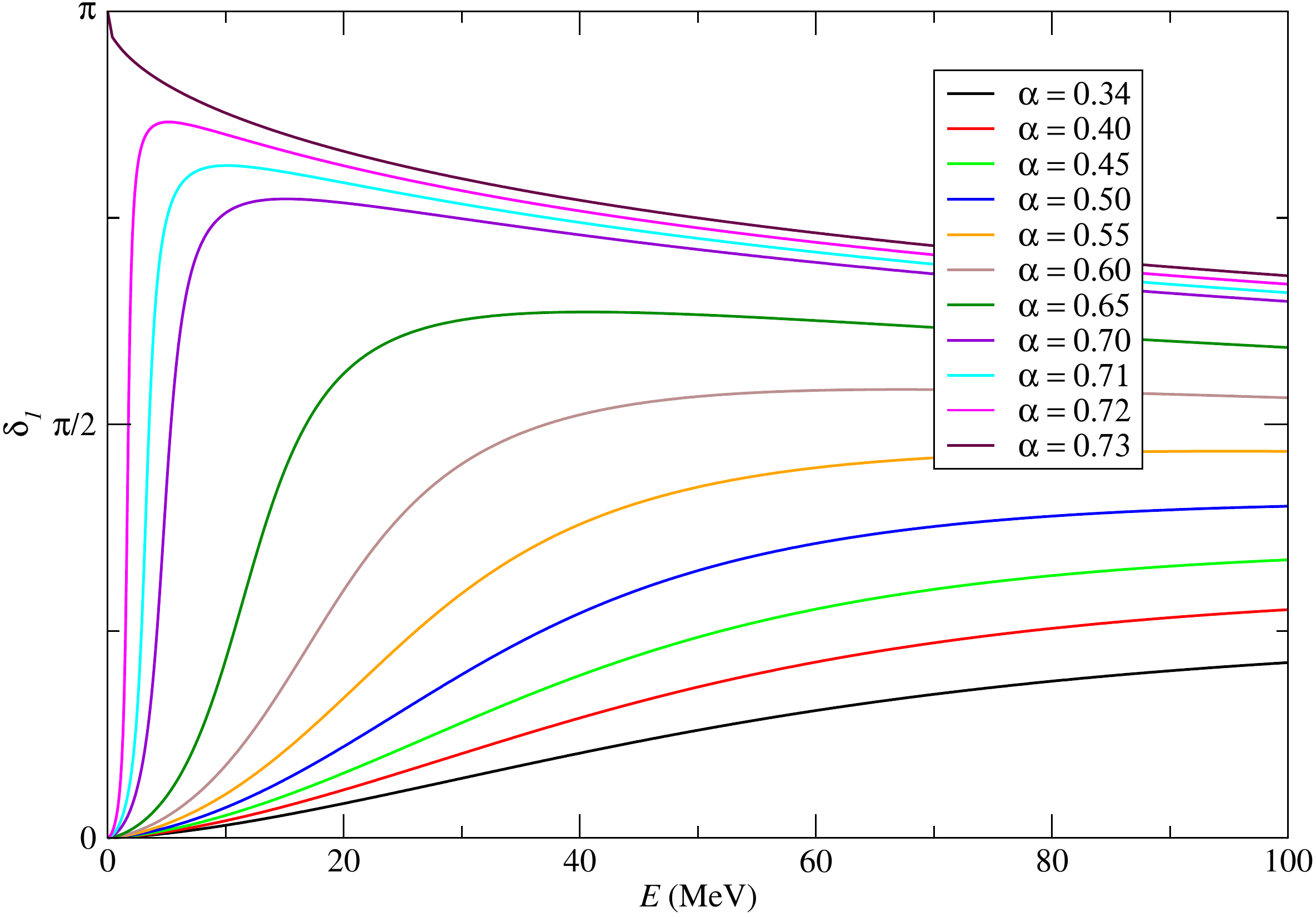}}
\caption{(Colour online.)
Phase shift $\delta_1$ as a function of the energy $E$ for different parameters for the potential. For illustration, we vary parameter $\alpha$ only while fixing $d = 0.45 \, \textrm{fm}$ at the value of the $(I = 0, j = 0)$ potential. Fixing $d$ and varying $\alpha$ produces comparable results.}
\label{fig:delta}
\end{figure}

Starting with angular momentum $l = 1$ we first search for clear resonance signals by making the potential more and more attractive. We increase the parameter $\alpha$, while keeping the parameter $d = 0.45 \, \textrm{fm}$ fixed, to preserve the scale of the potential. The corresponding results for the phase shift $\delta_1$ are shown in Fig.\ \ref{fig:delta}. Indeed, for $\alpha \gtapprox 0.65$ we find clear resonances with $\delta_1$ increasing from $0$ to $\approx \pi$. Then, for $\alpha = 0.72$, we find a bound state, since the phase shift $\delta_1$ starts at $\pi$ and decreases monotonically to $0$, when increasing the energy $E$. However, from these phase shifts it is not clear, for which values of $\alpha$ a resonance exists or not, i.e.\ it is not possible to say, whether there is a resonance for e.g.\ $\alpha \approx 0.50$ or even for the physical $\alpha = 0.34$.

\begin{figure}[t!]
\centerline{%
\includegraphics[width=1.05\columnwidth]{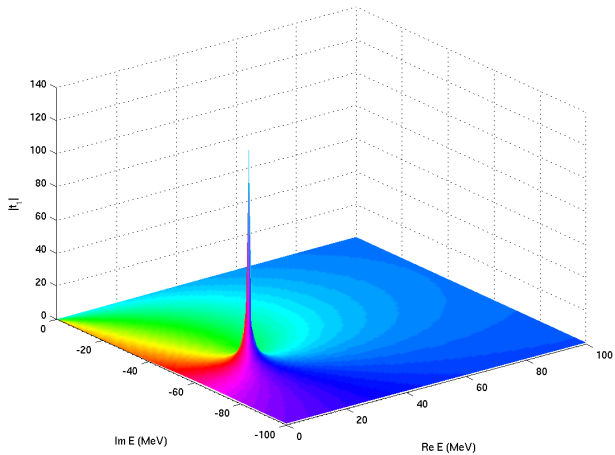}}
\caption{(Colour online.)
$\mbox{T}$ matrix eigenvalue $t_1$ as a function of the complex energy $E$ for the $(I = 0, j = 0)$ potential ($\alpha = 0.34$, $d = 0.45 \, \textrm{fm}$). Along the vertical axis we show the norm $|t_1|$, while the phase $\textrm{arg}(t_l)$ corresponds to different colours.
\label{fig:complex}}
\end{figure}

Thus, we search directly for poles of the $\mbox{T}$ matrix eigenvalues $t_l$. With this technique we clearly find a pole for angular momentum $l = 1$ and physical values of the parameters, $\alpha = 0.34$ and  $d =0.45 \, \textrm{fm}$. We show this pole in Fig.\ \ref{fig:complex} by plotting $t_1$ as a function of the complex energy $E$. The pole is clearly visible as a sharp peak.

\begin{figure}[t!]
\centerline{%
\includegraphics[width=0.95\columnwidth]{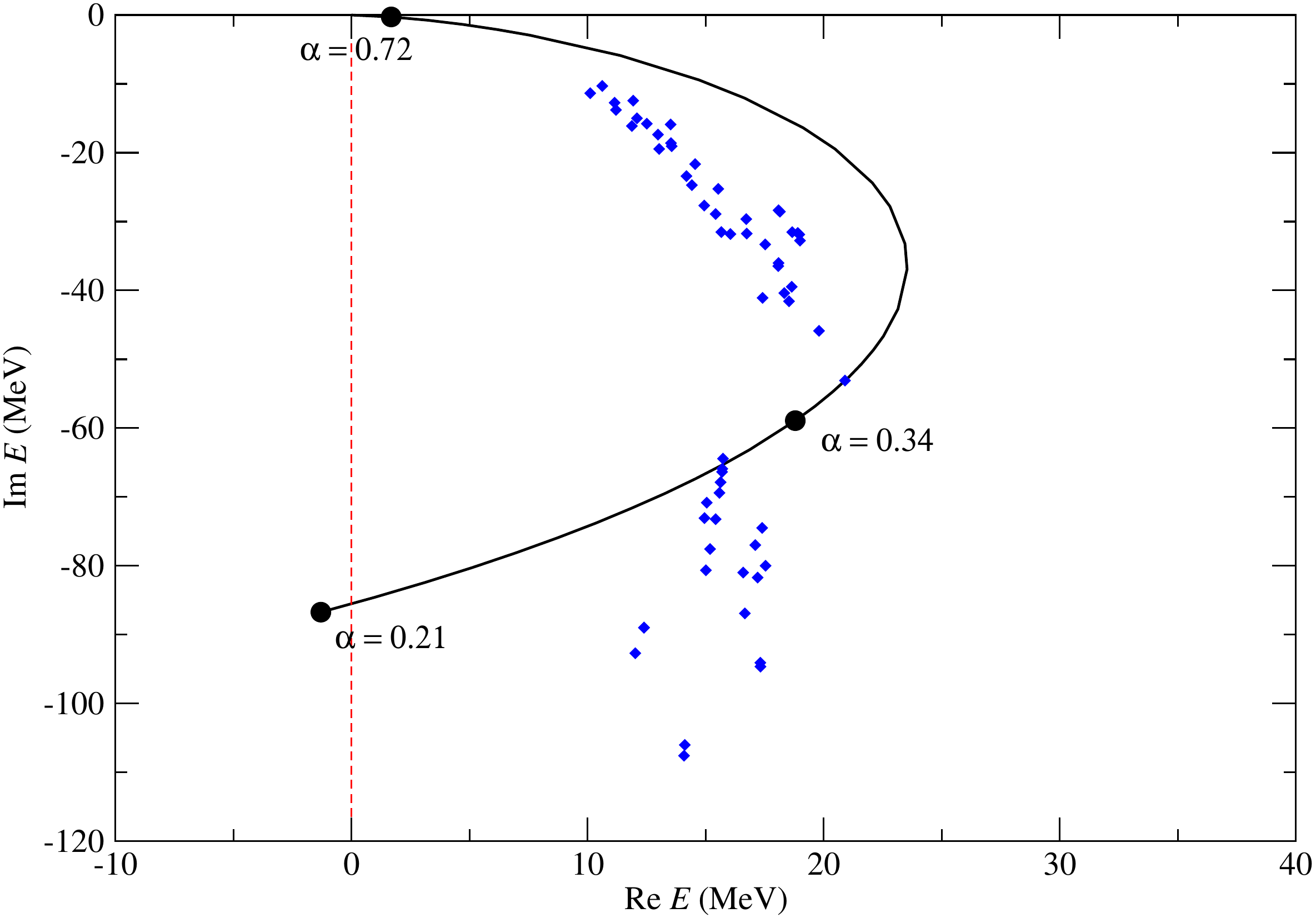}}
\caption{(Colour online) 
Locations for the  pole of the eigenvalue $t_1$ of the $\mbox{T}$ matrix  in the complex plane $(\textrm{Re}(E) , \textrm{Im}(E))$. We illustrate with a cloud of diamond points the computation of the systematic error of the  $\alpha$ and $d$ parameters of the  $(I = 0, j = 0)$ potential, utilizing the technique of Ref.  \cite{Bicudo:2015vta}.
We also depict (solid line) the trajectory of the pole corresponding to a variation of the potential parameters, varying $\alpha$ for $d = 0.45 \, \textrm{fm}$.
\label{fig:traj}}
\end{figure}

To understand the dependence of the resonance pole on the shape of the potential, we again scan different values of the parameter $\alpha$ and determine each time the pole of the eigenvalue $t_1$ of the $\mbox{T}$ matrix. We show the trajectory of the pole corresponding to a variation of $\alpha$ in the complex plane $(\textrm{Re}(E) , \textrm{Im}(E))$ in Fig.\ \ref{fig:traj}. Indeed, starting with $\alpha = 0.21$ we find a pole. This confirms our prediction of a resonance for angular momentum $l = 1$ and physical values of the parameters, $\alpha = 0.34$ and  $d =0.45 \, \textrm{fm}$.

Finally we perform a detailed statistical and systematic error analysis of the pole of $t_1$ and the corresponding values $(\textrm{Re}(E) , \textrm{Im}(E))$.  We use the same analysis method as for our previous study of the bound state for $l = 0$, cf.\ \cite{Bicudo:2015vta}. To parameterize the lattice QCD data for the potentials, $V^{\textrm{lat}}(r)$, discussed in Section \ref{sec:lattice}, we perform uncorrelated $\chi^2$ minimizing fits with the ansatz \eqref{eq:potential}. To this end we minimize the expression
\begin{equation}
\chi^2 = \sum_{r=r_{\textrm{min}},..., r_{\textrm{max}}}\left( \frac{V(r)-V^{\textrm{lat}}(r)}{\Delta V^{\textrm{lat}}(r)}  \right)^2
\label{eq:chisquared}
\end{equation}
with respect to the parameters $\alpha$, $d$ and $V_0$ ($\Delta V^{\textrm{lat}}(r)$ denote the corresponding statistical errors). To quantify systematic errors, we perform a large number of fits, where we vary the following parameters:
\begin{itemize}
\item The range of temporal separations $t_{\textrm{min}}\leq t\leq t_{\textrm{max}}$ of the correlation function \eqref{eq:corr_fnc}, where $V^{\textrm{lat}}(r)$ is read off, according to
\begin{itemize}
\item $t_{\textrm{max}} - t_{\textrm{min}} \geq a$,
\item $4a \leq t_{\textrm{min}}$, $t_{\textrm{max}} \leq 9a$
\end{itemize}
($a \approx 0.079 \, \textrm{fm}$ is the lattice spacing). 
\item The range of spatial $\bar b \bar b$ separations $r_{\textrm{min}}\leq r\leq r_{\textrm{max}}$ considered in the $\chi^2$ minimizing fits to determine the parameters $\alpha$, $d$ and $V_0$ according to
\begin{itemize}
\item $r_{\textrm{min}} \in \{ 2a, 3a \}$,
\item $r_{\textrm{max}} \in \{ 8a, 9a, 10a \}$.
\end{itemize}
\end{itemize}
We obtain a large number of different, but similar potential parameterizations $V(r)$ characterized by sets of values for $\alpha$, $d$ and $V_0$. For each potential parameterization we determine the position of the pole of $t_1$, i.e.\ $(\textrm{Re}(E) , \textrm{Im}(E))$ as discussed above and shown as a cloud of blue points in Fig.\ \ref{fig:traj}. For both $\textrm{Re}(E)$ and $\textrm{Im}(E)$ we construct a distribution by considering all corresponding results weighted by $\exp(-\chi^2/\textrm{dof})$ with $\chi^2$ from Eq.\ \eqref{eq:chisquared}. The central values of $\textrm{Re}(E)$ and $\textrm{Im}(E)$ are then defined as the medians of the corresponding distributions and the lower/upper systematic uncertainties are given by the differences of the 16th/84th percentiles to the medians. To also include statistical errors, we compute the jackknife errors of the medians of $\textrm{Re}(E)$ and $\textrm{Im}(E)$ and add them in quadrature to the corresponding systematic uncertainties. With our combined statistical and systematic error analysis we find a resonance energy $\textrm{Re}(E) = 17^{+4}_{-4} \, \textrm{MeV}$ and a decay width $\Gamma = -2 \textrm{Im}(E) = 112^{+90}_{-103} \, \textrm{MeV}$. Using the Pauli principle and considering the symmetry of the quarks with respect to colour, flavour, spin and their spatial wave function one can determine the quantum numbers of the resonance, which are $I(J^P) = 0(1^-)$. The resonance will decay into two $B$ mesons and, hence, its mass is $m = 2 M + \textrm{Re}(E) = 10 \, 576^{+4}_{-4} \, \textrm{MeV}$.

Note that there is also an uncertainty associated with the lattice spacing, $a = 0.079(3) \, \textrm{fm}$ (cf.\ Ref.\ \cite{Baron:2009wt} for details), which has not been taken into account so far. We have investigated the impact of this uncertainty on our final results for the resonance energy $\textrm{Re}(E)$ and the decay width $\Gamma$. We have found that both quantities exhibit only a mild dependence on the lattice spacing $a$ and the propagation of the uncertainty of $a$ has a negligible effect on the results for $\textrm{Re}(E)$ and $\Gamma$ quoted above within the current combined systematic and statistical errors.

In what concerns angular momenta $l \neq 1$, we find no clear signal for a resonance pole (except for the bound state pole for $l = 0$). We also find no poles for any $l$ in the less attractive case of $(I = 1 , j = 1)$.


\section{Conclusions and outlook \label{sec:conclusion}}

As a case study for the investigation of resonances above the $B B$ meson pair threshold, we have explored the $u d \bar{b} \bar{b}$ four-quark system. We have utilized lattice QCD potentials computed for two static antiquarks in the presence of two light quarks, the Born-Oppenheimer approximation and the emergent wave method for the $B B$ system. First we have computed scattering phase shifts. Then we have performed the analytic continuation of the $\mbox{S}$ matrix and the $\mbox{T}$ matrix to the second Riemann sheet and have searched for poles as signals of resonances.

From these results we have predicted a new resonance, with quantum numbers $I(J^P) = 0(1^-)$. Performing a careful statistical and systematic error analysis has led to a resonance mass $m = 10 \, 576^{+4}_{-4} \, \textrm{MeV}$ and a decay width $\Gamma = 112^{+90}_{-103} \, \textrm{MeV}$.


\begin{acknowledgements}

We acknowledge useful conversations with K.~Cichy.

P.B.\ acknowledges the support of CFTP (grant FCT UID/FIS/00777/2013) and is thankful for hospitality at the Institute of Theoretical Physics of Johann Wolfgang Goethe-University Frankfurt am Main. M.C.\ acknowledges the support of CFTP and the FCT contract SFRH/BPD/73140/2010. M.W.\ acknowledges support by the Emmy Noether Programme of the DFG (German Research Foundation), grant WA 3000/1-1.

This work was supported in part by the Helmholtz International Center for FAIR within the framework of the LOEWE program launched by the State of Hesse.

Calculations on the LOEWE-CSC and on the on the FUCHS-CSC high-performance computer of the Frankfurt University were conducted for this research. We would like to thank HPC-Hessen, funded by the State Ministry of Higher Education, Research and the Arts, for programming advice.

\end{acknowledgements}

\bibliographystyle{apsrev4-1}
\bibliography{literature.bib}

\end{document}